\documentclass[12pt]{article}
\usepackage{latexsym}
\usepackage{amsmath}
\usepackage{amssymb}
\usepackage{epsfig,graphics}
\usepackage{booktabs}
\usepackage{multirow}

\newcommand{\be}{\begin{equation}}
\newcommand{\ee}{\end{equation}}

\begin{document}

\begin{titlepage}

\vspace*{0.6in}
 
\begin{center}
{\large\bf SO(2N) and SU(N) gauge theories in 2+1 dimensions}\\
\vspace*{0.85in}
{Francis Bursa$^{a}$, Richard Lau$^{b}$ and Michael Teper$^{b}$\\
\vspace*{.2in}
$^{a}$Physics Department, Swansea  University, Swansea SA2 8PP, UK \\
\vspace*{.1in}
$^{b}$Rudolf Peierls Centre for Theoretical Physics, University of Oxford,\\
1 Keble Road, Oxford OX1 3NP, UK
}
\end{center}

\vspace*{0.4in}

\begin{center}
{\bf Abstract}
\end{center}

We perform an exploratory  investigation of how rapidly the
physics of SO($2N$) gauge theories approaches its $N=\infty$ limit.
This question has recently become topical because SO($2N$) gauge
theories are orbifold equivalent to SU($N$) gauge theories, but do not
have a finite chemical potential sign problem. We consider only the
pure gauge theory and, because of the inconvenient location of the
lattice strong-to-weak coupling 'bulk' transition in 3+1 dimensions, 
we largely confine our numerical calculations to 2+1 dimensions.  
We discuss analytic expectations in both $D=2+1$ and  $D=3+1$,
show that the SO(6) and SU(4) spectra do indeed appear to be the same, and show
that a number of mass ratios do indeed appear to agree in the
large-$N$ limit. In particular SO(6) and SU(3) gauge theories are
quite similar except for the values of the string tension and
coupling, both of which differences can be readily understood.

\vspace*{0.95in}

\leftline{{\it E-mail:} f.bursa@swansea.ac.uk,  r.lau1@physics.ox.ac.uk, m.teper1@physics.ox.ac.uk}

\end{titlepage}

\setcounter{page}{1}
\newpage
\pagestyle{plain}

\tableofcontents

\section{Introduction}
\label{section_intro}

SO($2N$) gauge theories are of topical interest because they do not
suffer a finite chemical potential sign problem
\cite{AC_mu},
are orbifold equivalent to SU($N$) gauge theories 
and share with the latter  a common large $N$ limit
in their common sector of states (see 
\cite{AC_N}
and references therein).
Thus if SU(3) and, say, SO(6) are both close to $N=\infty$, then
the finite baryon density phase diagram in QCD might be 
illuminated by lattice Monte Carlo calculations of, say, SO(6)
\cite{AC_mu}.

This has motivated us to study the pure gauge theories, 
as a first step. In this case much is known about 
SU($N$) both in D=3+1
\cite{G_D4N,GK_D4N}
and in D=2+1
\cite{G_D3N},
and our calculations have therefore focused upon SO($N$). As shown
below, if one uses the standard plaquette action then the strong-to-weak 
coupling `bulk' phase transition in D=3+1 SO($N$) gauge theories
occurs at such a small value of the lattice spacing $a$, when  $N$ is not large,
that it becomes prohibitively expensive to perform weak coupling
calculations in volumes that are large enough to be in the confining
phase. (This has long been known in the extreme case of SO(3). For a recent study see
\cite{PDF_SU2adj}.) 
To deal with this problem, we are currently exploring improved actions in 
the hope that the bulk transition may be shifted to stronger
coupling. In D=2+1 on the other hand, the strong-to-weak coupling transition 
provides much less of an obstacle and so most of the lattice calculations in
the present paper will deal with SO($N$) gauge theories in 2+1 dimensions.

There are of course additional reasons for being interested in SO($N$)
gauge theories. For example, an SO($N$) gauge theory can have
exactly the same Lie algebra as some SU($N^\prime$) theory. This is the
case for SO(3) and SU(2) and also for SO(6) and SU(4). One would expect
the two theories in each pair to have the same spectrum in the continuum
limit, assuming the global properties of the group do not play a role
in the dynamics. It would be nice to check this expectation, and we
shall provide evidence later on in this paper that this is indeed the
case for SU(4) and SO(6). This will enable us to make an approximate
prediction for the $N$-dependence of SO($N$) gauge theories for 
$N\geq 6$ in both 2+1 and 3+1 dimensions, using the known properties
of the SU(4) gauge theory.

Although our calculations are still at an early stage, the results
we have obtained provide useful information on the above questions
and we will present these below. We will also discuss
in detail how to compare physical quantities in SU($N$) and SO($2N$) gauge theories.
We will present results for the string tension, $\sigma$, the 
lightest two $J^P=0^+$ scalar states, the lightest $J^P=2^+$  tensor, the
deconfining temperature $T_c$, and the coupling $g^2$, which in D=2+1
has dimensions of mass. We do so for SO(4), SO(6), SO(8) and SO(12)
gauge theories, and use these results to test the large-$N$
equivalence with SU($N$) and to determine the rate of approach to that
limit.  

Since our primary focus here is on the SO($2N$) and SU($N$) equivalence, 
we do not discuss SO(2N+1) gauge theories, which differ from SU($2N$)
in that they lack the (useful) $Z_2$ center symmetry of the latter. We
will leave our detailed comparison of SO(3) and SU(2) to a separate
paper and will include our ongoing detailed study of odd $N$ to a future 
publication. We merely note here that at first glance the physics of
SO(2N+1) gauge theories appears to be entirely continuous with that
of neighbouring SO($2N$) gauge theories.

In the next section we review some expectations about the large-$N$ 
limit. We also discuss how precisely the calculated physics of SO(6) and SU(4)
gauge theories is to be identified. (And separately the case of SU(2) and
SO(3).) We note that this equality provides approximate predictions
for the $N$-dependence of SO($N$) gauge theories in both D=2+1 and
D=3+1. The following Section contains our calculations. We outline the
lattice calculation and then locate the strong-to-weak coupling  
transition in both D=3+1 and D=2+1. We show that in the former case
it is only for $N\geq 16$ that one can obtain useful weak-coupling physics on
reasonably sized lattices. Focusing on D=2+1 we provide a detailed
calculation for SO(6) and compare the continuum extrapolation to
SU(4). We then extrapolate to the continuum our calculations for other
values of $N$. Here our calculations are currently much more limited
and we need to justify the reliability of these extrapolations using
what we find in SO(6).
 
It is worth listing some of the ways in which our work in progress
\cite{fbrlmt_inprog}
will improve upon the D=2+1 results presented here. First, we will perform
calculations at smaller $a$ so as to reduce the systematic error on
our continuum extrapolations. We will also include $P=-$ states.
(Recall that in D=2+1 $J^\pm$ states are degenerate except 
possibly for $J=0$.) We will include $J=1$ states as well as $J=0,2$. 
All this will provide a larger spectrum of states, calculated with
greater precision. We also intend to perform the calculations of $T_c$
much more accurately using standard reweighting methods that
have been used in the case of SU($N$) (see e.g.
\cite{Tc_D4N,Tc_D3N}).
We will include SO(2N+1) gauge theories in all these studies. Finally,
for a useful comparison it may also prove necessary to repeat the
corresponding SU($N$) calculations with greater accuracy than that
currently available.  

More speculatively we hope that  our lattice action improvement will
enable us to obtain SO($N$) continuum physics in D=3+1 for the modest values 
of $N$ where finite chemical potential calculations might conceivably 
be performed. In addition it would be of interest to study the
spinorial representations  particularly in the context of the
corresponding flux tubes and string tensions.

\section{Expectations}
\label{section_expectations}

\subsection{general remarks}
\label{subsection_general}

As is well known, the analysis of diagrams to all orders tells us
\cite{tHooft_N}
that the large $N$ limit of SU($N$) gauge theories is achieved by keeping $g^2N$ 
fixed and  that the leading correction in the pure gauge theory is $O(1/N^2)$.  
A parallel analysis for SO($N$) gauge theories tells us
\cite{AC_N}
that $g^2N$ should be kept fixed but that the leading correction is  $O(1/N)$.
Moreover the $N=\infty$ limit of the two theories is the same if we choose
the SO($N$) value of $g^2$ to be twice the SU($N$) value, 
\begin{equation}
\left.g^2\right|_{SO(N)}
\stackrel{N\to\infty}{=}
2 \times \left.g^2\right|_{SU(N)}
\label{eqn_g}
\end{equation}
or equivalently if we match
SO($2N$) and SU($N$) theories at the same  coupling. There
exists a corresponding orbifold equivalence (see references in
\cite{AC_N})
but for our limited purposes it would not add anything to this large-$N$
analysis.

Of course SO($N$) gauge theories have trivial charge conjugation
properties and therefore the comparison with SU($N$) is only in the
$C=+$ sector. The two groups also differ qualitatively in their symmetry
properties: although SU($N^\prime$) and SO($N$)
gauge theories may be equivalent at the level of the Lie algebra,
the global properties differ. For example, SU(4) has a $Z_4$ center
while SO(6) has only $Z_2$ while in the pair SU(2) and SO(3) the
former has a $Z_2$ center while the latter has a trivial center.
Large fields may be sensitive to the center and it is therefore
interesting to test the expectations of the diagrammatic equivalences
at the non-perturbative level using lattice Monte Carlo techniques.

Matching physical quantities in  SO($N$) and  SU($N^\prime$) gauge
theories is straightforward for colour singlet quantities, such as
`glueball' masses. For flux tubes and string tensions, however, 
one needs to be more careful. Suppose one considers a flux tube
that wraps around a spatial torus of length $l$. For $l$ large the
calculated energy gives the string tension via $E \simeq \sigma l$.
In SU($N$) there are a variety of stable flux tubes labelled by the
value $k=1,2,..,N/2$ of their 
${\cal{N}}$-ality, and other unstable flux tubes,
such as the adjoint flux tube, carrying flux in various
representations. In the case of SU(2) and SO(3), it is well known that
the latter is equivalent to the former in the adjoint representation.
Thus SO(3) flux tubes correspond to SU(2) flux tubes that carry
adjoint flux, which indeed have $C=+$. The latter are of course 
unstable, and can decay into glueballs,
but this is consistent with the fact that SO(3) does not possess a
non-trivial center which would prevent the mixing of a winding flux
tube operator with contractible operators that project onto glueball states.
Thus the $\sigma$ extracted in SO(3) corresponds to the adjoint string 
tension in SU(2). Since we are interested in SO(2$N$), we will
not pursue the SO(3) $\sim$ SU(2) correspondence any further here
\cite{fbrlmt_so3}.
The SU(4)$\sim$SO(6) correspondence is however relevant and that will
be discussed below. More generally we note that the $Z_2$ center of
SO($2N$) ensures that these theories have stable flux tubes just like
SU($N$). Of course for large $N$ we may expect, by continuity, that
SO($2N+1$) theories also have stable flux tubes, but if so (and our
preliminary calculations indicate that this is indeed the case) then
it will be enforced by dynamics rather than a non-trivial center symmetry.

\subsection{SO(6), SU(4) and SO($2N$)}
\label{subsection_so6su4so2N}

As is well known, SU(4) and SO(6) have the same Lie algebra, so one
expects that the $C=+$ glueball spectra will be identical. 
Now we recall that in SU(4)
\begin{equation}
\underline{4} \otimes \underline{4}  =  \underline{6} \oplus \underline{10}
\label{eqn_44to6}
\end{equation}
(see e.g.
\cite{Georgi_Lie})
where the $\underline{6}$ corresponds to the $k=2$ antisymmetric
representation (which indeed is $C=+$ for SU(4)) and maps to the fundamental
$\underline{6}$ of SO(6). Thus in the
equivalence with SO(6) we are to think of $k=2A$ operators in SU(4) and  
the SO(6) string tension should be compared to the  $k=2A$
string tension in SU(4). In terms of the fundamental SU(4) string tension
this has values
\begin{equation}
\frac{\sigma_{2A}}{\sigma_f}
= 
\begin{cases}
1.355 \pm 0.009 & D=2+1\\
1.370  \pm 0.020 & D=3+1
\end{cases}
\label{eqn_2AtoF}
\end{equation}
in $D=2+1$
\cite{K_D3N}
and $D=3+1$
\cite{GK_D4N}.
This implies that we should compare mass ratios as:
\begin{equation}
\left.\frac{M_G}{\surd\sigma}\right|_{so6}
=
\left.\frac{M_G}{\surd\sigma_{2A}}\right|_{su4}
\label{eqn_Kso6su4}
\end{equation}

\begin{figure}[htb]
\begin	{center}
\leavevmode
\input	{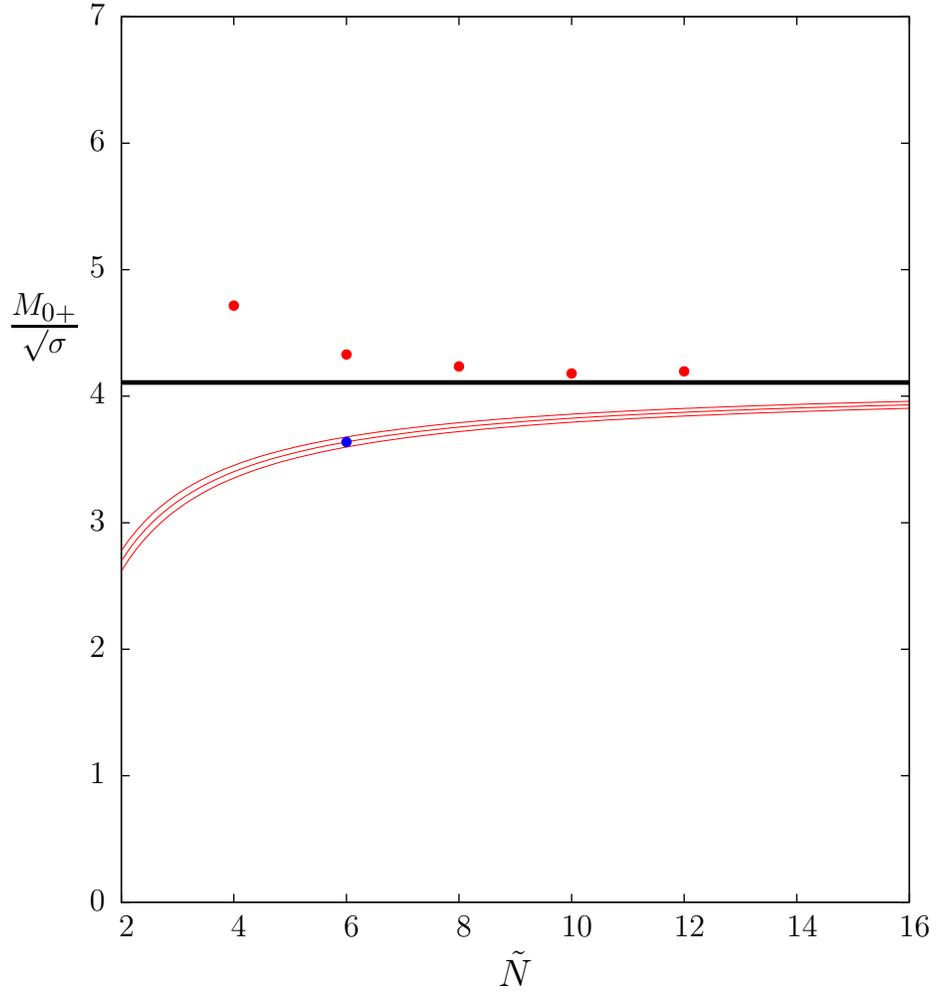}
\end	{center}
\caption{Mass of lightest glueball in units of  the string tension
  $\sigma$. Curves are predictions (with error band) for SO($\tilde{N}$) 
assuming just the leading $O(1/\tilde{N})$  correction and equality of spectra
for SO(6) and SU(4) and SO($\infty$) and SU($\infty$) as described in
the text. Red points are SU($N$) values at $N= \tilde{N}/2$.}
\label{fig_M0K_D3pred}
\end{figure}

For example, consider the lightest scalar glueball in D=2+1 SU(4). Using the
known value of the ratio in SU(4) 
\cite{G_D3N}
we obtain the corresponding ratio
in SO(6): 
\begin{equation}
\left.\frac{M_G}{\surd\sigma}\right|_{so6}
=
\left.\frac{M_G}{\surd\sigma_f}\right|_{su4}
\times
\left.\surd\left\{\frac{\sigma_f}{\sigma_{2A}}\right\}\right|_{su4}
=
\frac{4.235(25)}{1.164(4)}
=
3.638(25)
\label{eqn_M0Kso6su4}
\end{equation}
We also expect that at $N=\infty$,
\begin{equation}
\left.\frac{M_G}{\surd\sigma}\right|_{so(\infty)}
=
\left.\frac{M_G}{\surd\sigma_f}\right|_{su(\infty)}
=
4.108(20)
\label{eqn_M0KNinf}
\end{equation}
If we now assume that the leading $O(1/N)$ correction dominates for
$N\geq 6$, i.e.
\begin{equation}
\left.\frac{M_G}{\surd\sigma}\right|_{so(N)}
\simeq
\left.\frac{M_G}{\surd\sigma}\right|_{so(\infty)}
+
\frac{c}{N} \quad ; \  N\geq 6
\label{eqn_M0KN}
\end{equation}
we can use eqns(\ref{eqn_M0KNinf},\ref{eqn_M0Kso6su4}) to determine
the coefficient $c$ in eqn(\ref{eqn_M0KN}) and hence the ratio for all
values of $N\geq 6$. We display this prediction in
Fig.~\ref{fig_M0K_D3pred},
where we also display the SU($N$) values of the ratio
\cite{G_D3N}, 
and the value
predicted for SO(6) in  eqn(\ref{eqn_M0Kso6su4}). We see that this
ratio approaches the $N=\infty$ limit from opposite sides for SO($N$)
and SU($N$). This is driven by the fact that in the
denominator of this ratio we have 
$\surd\sigma|_{so6} = \surd\sigma_{2A}|_{su4} \simeq  1.164 \surd\sigma|_{su4}$.
The important corollary is that even if our assumption of the
dominance of the $O(1/N)$
correction is inaccurate, the qualitative behaviour shown in 
Fig.~\ref{fig_M0K_D3pred} is almost certain to survive. 

We can obviously extend the above argument to any glueball mass ratio: 
assume the dominance of the leading $O(1/N)$ correction for $N\geq 6$, 
then use existing SU(4) and SU($\infty$) results
\cite{G_D3N,G_D4N,GK_D4N,K_D3N}
to fix the mass ratio for SO(6) and SO($\infty$), and hence predict
the ratio for all SO($N\geq 6$). It is clear that 
the difference between SU($N$) and SO($N$) for such glueball
mass ratios is certain to be far more modest than for 
$M_G/\surd\sigma$. For example, $m_{2+}/m_{0+}$ 
and  $m_{0+\star}/m_{0+}$  change by less than $1\%$ when
we go from SU(4) to SU($\infty$) 
\cite{G_D3N,GK_D4N}
and hence also when we go from
SO(6) to SO($\infty$). Short of some fine-tuned cancellation
between the $O(1/N)$ and $O(1/N^2)$ corrections for SO(6), the near
constancy of these mass ratios for SO($N\geq 6)$ is thus more-or-less
guaranteed. The same comment applies to the location of the 
deconfining temperature $T_c$ if expressed in units of the mass gap. 
Irrespective of these arguments, the fact that such energy ratios
are known to be very similar for SU(3) and SU(4) immediately implies
the same for SU(3) and SO(6), i.e. in the case of particular interest to 
the finite chemical potential problem
\cite{AC_mu,AC_N}. 
(We focus here on the pairing of SU(3) and SO(6) because of the large-$N$ 
orbifold equivalence of SU($N$) and SO($2N$) gauge theories. However 
we should bear in mind that other SO($N$) theories that are
sufficiently `close' to SU(3) may be useful -- either because they are
less expensive, e.g. SO(4), or because they have a less severe bulk transition
problem, e.g. SO(12) in D=3+1.)

In D=2+1 the coupling $g^2$ has dimensions of mass, and so we can consider
ratios of $\mu/g^2N$ for some physical mass $\mu$, and ask how this ratio
approaches the $N=\infty$ limit where one expects that $g^2|_{soN}=2
g^2|_{suN}$
\cite{}.
 Consider SO(6). The large $N$ expectation 
\cite{tHooft_N,AC_N}
would be that
\begin{equation} 
g^2|_{so6}=2 g^2|_{su6} = 2 \times \frac{2}{3} g^2|_{su4} =
\frac{4}{3} g^2|_{su4}
\label{eqn_gso6exp}
\end{equation}
 Now we know that
the SO(6) action is equivalent to working in SU(4) with the fields in the
$k=2A$ representation. One can think of a mixed SU(4) lattice
plaquette action 
\begin{equation}
S = \beta_f \sum_p \{1-\frac{1}{N_f}ReTr_f u_p\}
+
\beta_{2A} \sum_p \{1-\frac{1}{N_{2A}}Tr_{2A} u_p\}
\label{eqn_Smixed}
\end{equation}
where 
\begin{equation}
\beta_f = 2N_f/g^2_f \quad ; \quad \beta_{2A}=2N_{2A}/g^2_{2A}
\label{eqn_beta}
\end{equation}
just like a more conventional mixed fundamental-adjoint action. We
have added here a subscript $f$ to the usual (fundamental) $g^2$ for
clarity. For SU(4), the sizes of the representations are
$N_f=4, \ N_{2A}=6$. Using 
\begin{equation}
Tr_{2A} u_p = \frac{1}{2}\left\{ (Tr_fu_p)^2- Tr_fu^2_p \right\}
\label{eqn_trace2A}
\end{equation}
and performing a weak coupling expansion one readily sees that
\begin{equation}
g^2|_{so6} = g^2_{2A}|_{su4} = 2 g^2_f|_{su4}
\label{eqn_gso6}
\end{equation}
This differs from the large-$N$ expectation in eqn(\ref{eqn_gso6exp})
by a factor of 1.5, implying that here, just as with the string
tension, there are substantial finite $N$ corrections.

We can conclude from the above general arguments that the comparison between
SU(3) and SO(6) is as follows: ratios of glueball masses and $T_c$ are
very similar, while ratios involving the string tension and the
coupling will differ at the $15-20 \%$ level.

Of course, all the above assumes that the different global properties
of the SU(4) and SO(6) groups plays no important role in the details
of the spectrum. This is something that we shall explicitly check for the
lightest masses below.

\subsection{SO(3) and SU(2)}
\label{subsection_so3su2}

Although we do not study SO(2N+1) theories in this paper, it is
relevant to note that, using the fact that SO(3) is the
adjoint of SU(2), one can further constrain the $N$-dependence of
SO($N$) gauge theories, using the known properties of SU(2) gauge
theories.  Together with the constraint from SO(6) and SU(4) this
allows us to fix both the $O(1/N)$ and $O(1/N^2)$ corrections
if we assume that these dominate down to SO(3). Care is needed
with the string tension, since the SU(2) adjoint string is unstable, but 
ratios involving glueball masses, $T_c$, and $g^2$ can be treated
straighforwardly by an obvious extension of the analysis in
Section~\ref{subsection_so6su4so2N}.

\section{Calculations in D=2+1}
\label{section_calculations}

\subsection{calculating on the lattice}
\label{subsection_lattice}

Our lattice field variables are SO($N$) matrices, $U_l$, residing on the links $l$
of the $L^2_s L_t$ lattice, whose spacing is $a$. 
(We will employ the same notation as used for unitary
matrices although here the matrices are of course real.) The Euclidean path
integral is $Z=\int {\cal{D}}U exp\{-S[U]\}$ and we use the standard plaquette action,
\begin{equation}
S = \beta \sum_p \left\{1-\frac{1}{N} Tr U_p\right\}  \quad ; \quad \beta=\frac{2N}{ag^2}
\label{eqn_S}
\end{equation}
where $U_p$ is the ordered product of link matrices around the
plaquette $p$. We update the fields using a natural extension
 to SO($N$) of the SU($N$) Cabibbo-Marinari algorithm. (The details
of this algorithm will be described elsewhere 
\cite{fbrlmt_inprog}.)

Our SO($N$) calculations closely parallel those in SU($N$), so we will be
very brief here and will refer the reader to other papers for
details.

The particle (`glueball') states can be labelled by parity $P=\pm$ and
spin $J$.  (Charge
conjugation is necessarily positive.) For $J\neq 0$ the  $P=\pm$
states are necessarily degenerate in $D=2+1$ 
\cite{G_D3N}
and in this exploratory
study we shall only calculate the masses of $P=+$ and $J=0,2$ states.
Here we will make the usual simplifying assumption that the states we
see have the lowest $J$ that contributes to the relevant square
lattice representation. (This is usually but not always the case
\cite{hmmt_D3G}.) 

Ground state masses $M$ are calculated from the asymptotic time
dependence of correlators, i.e.
\begin{equation}
<\phi(t) \phi(0)> \stackrel{t\to\infty}{\propto} e^{-Mt} 
\label{eqn_M}
\end{equation}
where $M$ is the mass of the lightest state with the quantum numbers
of the operator $\phi$. To calculate excited states as well one
calculates (cross)correlators of several operators and uses these as a
basis for a systematic variational calculation in $e^{-Ht_0}$ where $H$
is the Hamiltonian (corresponding to our lattice transfer matrix) and
$t_0$ is some convenient distance. To have good overlaps onto the
desired states, so that one can evaluate masses at values of $t$
where the signal has not yet disappeared into the statistical noise,
one uses blocked and smeared operators. (For details see e.g.
\cite{G_D3N,GK_D4N}.) 

To calculate the string tension $\sigma$ we use the above technique to
calculate the energy $E$ of the lightest flux tube that winds around one
of the periodic spatial tori. If the length $l=aL_s$ of the torus is
large then $E(l) \simeq \sigma l$ where $\sigma$ is the string
tension. There are of course corrections and we assume that
for our range of $l$ these are accurately incorporated in the simple
Nambu-Goto expression 
\begin{equation}
E(l)=\sigma l \left\{ 1 - \frac{\pi}{3\sigma l^2} \right\}^{1/2}
\label{eqn_NG}
\end{equation}
which is what we shall use to extract $\sigma$ from $E(l)$. (See e.g.
\cite{string_MC,Aharony}
and references therein.)

The operator we use is the Polyakov loop $l_p$, i.e. the product 
of link matrices along a minimal length curve that closes around the
spatial torus. (And blocked versions of this.) For even $N$, which is
the case of interest in this paper, the theory has a $Z_2$ symmetry
that ensures that $<l_p> = 0$ as long as the symmetry is not
spontaneously broken, and indeed that $<l_p \phi_G> = 0$ where
$\phi_G$ is any contractible loop (which is what one uses for
glueball operators). That is to say we have a stable flux tube state that winds 
around the torus. 

We can similarly consider Polyakov loop operators that wind around 
the temporal torus on our $L^2_sL_t$ lattice. Such a finite torus 
corresponds to a finite temperature $T=1/aL_t$ if we are in the
thermodynamic limit $L_s \gg L_t$. The Polyakov loop is the
contribution to the action of a single charged static source.
Just as above, the $Z_2$ symmetry ensures that  $<l_p> = 0$ i.e. that 
the free energy of the static source is infinite and we are in
the low temperature confining phase. As we decrease $L_t$ at some
temperature $T=T_c$ the $Z_2$ symmetry spontaneously breaks, we
have  $<l_p> \neq 0$  and 
we enter the deconfined phase where we have Debye screening and the
source has a finite free energy.

\subsection{bulk transition}
\label{subsection_bulk}

Lattice gauge theories generally show a (`bulk') transition between the
strong and weak  coupling regions where the natural expansion parameters
are $\beta \propto 1/g^2$ and  $1/\beta \propto g^2$ respectively.
Since the extrapolation to the continuum limit should be made within the
weak coupling region, it is important that the bulk transition 
should occur at a value
of $\beta$ where $a$ on the weak coupling side is not very small.
Otherwise prohibitively large lattices may be needed to ensure that
one is in the weak  coupling confining phase. 

For D=3+1 SU($N$) gauge theories it is known that the transition
is first order for $N\geq 5$ and is a cross-over for smaller $N$
\cite{G_D4N}. 
In D=2+1 it appears to
\cite{fbmt_bulk}
 be quite similar to the Gross-Witten transition in
D=1+1 
\cite{GW}
i.e. a cross-over for all $N<\infty$ developing into a
third-order transition at  $N=\infty$. The location in $D=3+1$
is such that on the weak coupling side we can readily go down to  
$a \sim 1/5T_c$ (taking advantage of the metastable region when the
transition is first order). In $D=2+1$, we can go to much larger $a$,
$a \sim 1/1.6T_c$.  So in these cases the bulk transition presents no significant
obstacle  to continuum extrapolations. On the other hand it has long
been known that for the $D=3+1$ SU(2) theory in the adjoint representation,
there is a bulk phase transition with a very small (and not precisely
known)  value of $a$ on the weak coupling side. (For a recent
discussion see
\cite{PDF_SU2adj}.)

\begin{figure}[htb]
\begin	{center}
\leavevmode
\input	{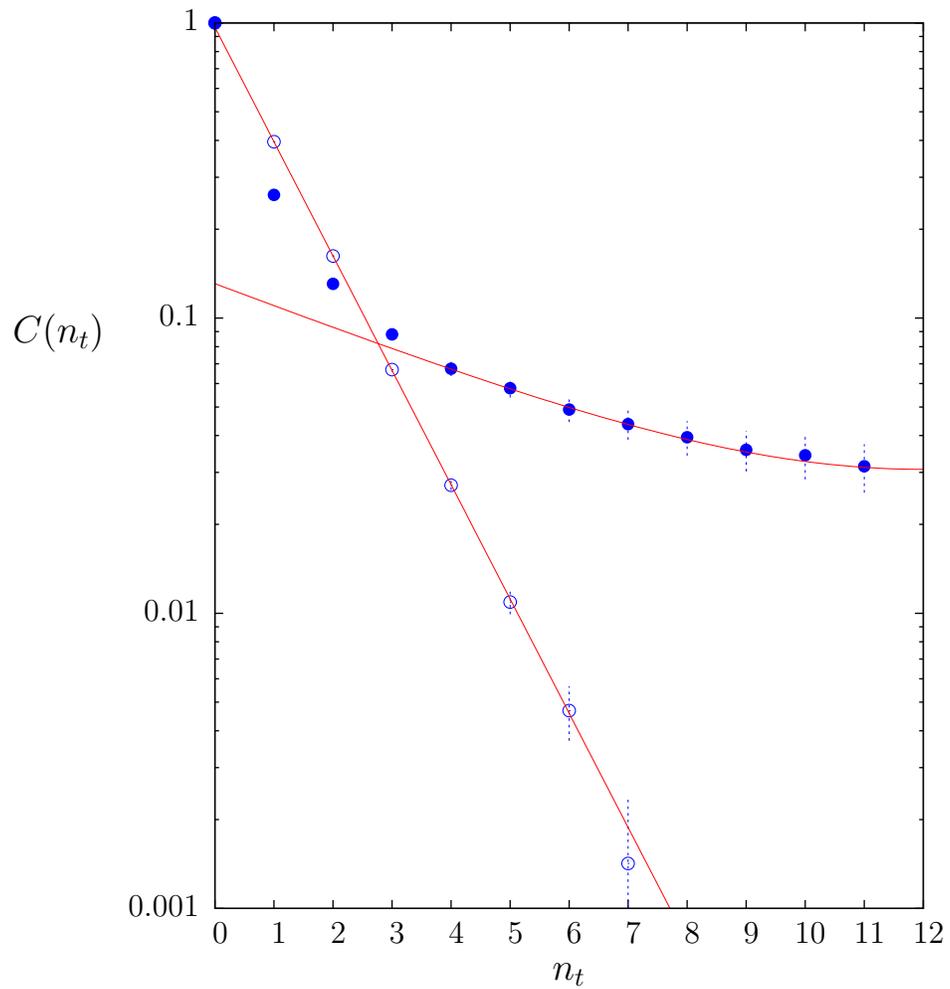}
\end	{center}
\caption {Correlation functions of the `lightest', $\circ$, and
`first excited', $\bullet$, $0^+$  glueball states in SO(4) on a
  $16^2 24$ lattice  at $\beta=9.3$. Single mass cosh fits are shown.}
\label{fig_bulk}
\end{figure}

Since adjoint SU(2) is the same as SO(3), this suggests that 
in $D=3+1$ the location of the bulk transition may be an obstacle
to accessing the continuum limit of SO($N$) gauge theories.
We will address this in more detail later on in the paper. Here we
turn to SO($N$) gauge theories in D=2+1. We have performed
scans in $\beta$ for various $N$ which show no sign of any first order
transition. However we do find a transition which is characterised by
a near-vanishing of a scalar glueball mass. The transition is in a
narrow range of $\beta$ and its location depends slightly on the
volume of the lattice. An example is shown in 
Fig~\ref{fig_bulk}. Here we show the correlation functions of the 
two lightest glueballs as obtained from our variational procedure that 
maximises $e^{-aH}$ over the basis of operators. The `lightest'
glueball is well fitted by a single cosh, showing that it has a very
good overlap onto our basis. It is the state that is continuous with
the lightest glueball masses away from the phase transition. The `first
excited state' , on the other hand, shows the presence of a very light
particle that only shows up at larger $n_t$ because it has a small
overlap onto our basis. The presence of this light particle is the
signal for the bulk transition. It is possible that we are seeing a
nearby critical point, which might indeed be a second order phase
transition at a nearby value of $\beta$. We have not investigated 
the nature of this transition or cross-over
any further except to list in Table~\ref{table_bulk} the values of
the 't Hooft coupling, $ag^2N=2N^2/\beta$ at which we have observed
it to occur. For comparison we show an estimate of the location for
SU($\infty$) theories
\cite{fbmt_bulk}.
We note that our results are roughly consistent with the 
naive orbifold expectation that $g^2_b$ for SO($2N$) and SU($N$)
lattice gauge theories should become the same as $N\to\infty$ i.e.
that the 't Hooft couplings should differ by a factor of 2.
We also show the corresponding values of the string
tension. We see that the transition occurs at a modest
value of $a$ in units of the string tension and so should present no 
significant obstacle to a continuum extrapolation.

\subsection{SO(6) and SU(4)}
\label{subsection_so6su4}

In the case of SO(6)  we have
performed calculations over an extended range of $a$, designed to
minimise any systematic error in performing the continuum
extrapolation, so as to make our comparison with existing results for
SU(4) reasonably reliable. The parameters of these calculations and
some of the physical quantities calculated are shown in Table~\ref{table_SO6}. 

As shown in Table~\ref{table_SO6}, we have  performed calculations
with various spatial volumes at $\beta=29$, in order to determine
how large a volume we need in order to avoid finite volume corrections
(within the statistical errors characteristic of all our calculations). 
We observe no corrections for $L_s\geq 16$, except possibly 
for the excited scalar glueball which appears to require $L_s\geq 20$.
In physical units these two lattice sizes correspond to  
$L_sa\surd\sigma \sim 3.4,\ 4.3$ respectively. We note that
all the other calculations in  Table~\ref{table_SO6} which are beyond
the bulk transition ($\beta_b \sim 19$)  have been chosen to satisfy 
the first bound, and
that the ones at the smallest values of $a$ are close to the second
bound. So we expect finite volume corrections to be small
in our continuum calculations.

\begin{figure}[htb]
\begin	{center}
\leavevmode
\input	{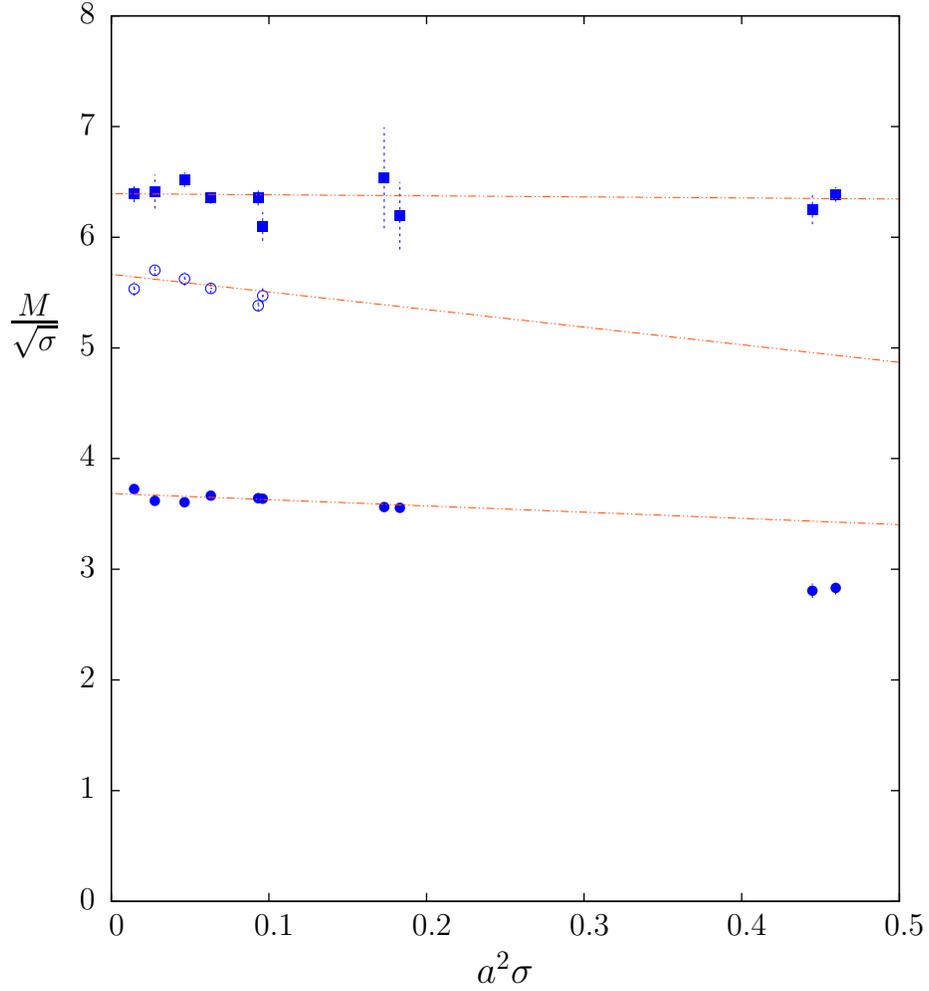}
\end	{center}
\caption {Masses of the ground state $0+$ ($\bullet$), first excited  $0+$ ($\circ$),
and ground state  $2+$ ($\blacksquare$)  in units of the string tension,
plotted versus the string tension, and with $O(a^2)$ continuum
extrapolations shown. All for SO(6) in D=2+1.} 
\label{fig_MK_so6}
\end{figure}

Taking ratios of glueball masses to the string tension, we can attempt
to extrapolate to the continuum limit using just a leading $O(a^2)$ 
lattice correction
\begin{equation}
\left.\frac{aM_G}{a\surd\sigma}\right|_a
=
\left.\frac{M_G}{\surd\sigma}\right|_a
=
\left.\frac{M_G}{\surd\sigma}\right|_{a=0}
+
c a^2\sigma .
\label{eqn_MKcont}
\end{equation}
In Fig.~\ref{fig_MK_so6} we plot this ratio for the lightest two
scalar glueballs and the lightest tensor glueball. (Note that the
light scalar associated with the bulk transition at $\beta \sim18$ 
is deliberately excluded.) We show linear continuum extrapolations of the form 
in eqn(\ref{eqn_MKcont}) and these seem reasonably well determined.
The resulting continuum mass ratios, obtained using values of $\beta$
beyond the bulk transition, are listed in
Table~\ref{table_E_SO6SU4}. We have also shown there the values
obtained from fits in which eqn(\ref{eqn_MKcont})  is supplemented
by an additional $O(a^4)$ correction. The difference between the pair
of fits provides an estimate of one of the systematic errors in our 
continuum extrapolations. We also show the deconfining temperature
whose calculation we leave to a later section. Finally, for comparison, 
we show the corresponding results for the SU(4) gauge theory
\cite{}.
The agreement is very good at the 2 standard deviation level.
This provides direct confirmation of the expected equivalence
of the SU(4) and SO(6) spectra, and of our identification of the 
SO(6) string tension with the $k=2A$ string tension of SU(4).

\begin{figure}[htb]
\begin	{center}
\leavevmode
\input	{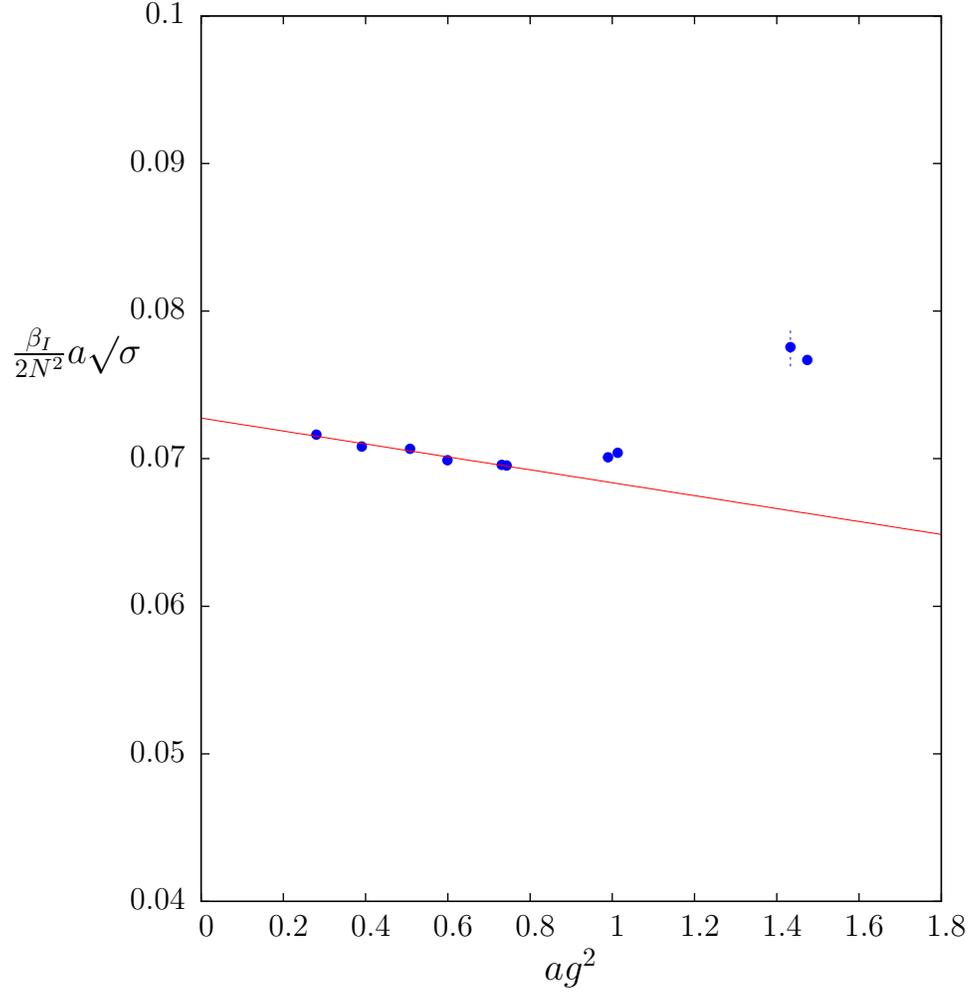}
\end	{center}
\caption {String tension in units of $g^2N$ (using $\beta_I=2N/ag^2$)
with an $O(ag^2)$ extrapolation to the continuum limit. For SO(6) in D=2+1.} 
\label{fig_Kg_so6}
\end{figure}

There remains one major prediction to test: the relationship between
the SO(6) and SU(4) couplings given in eqn(\ref{eqn_gso6}).
This can be done by calculating $\surd\sigma/g^2$ in SO(6) and
comparing to the SU(4) value of $\surd\sigma_{2A}/g^2$: the former
is then predicted by  eqn(\ref{eqn_gso6}) to be one-half of the latter. 
To obtain the SO(6) value of this ratio we consider the continuum
extrapolation
\begin{equation}
\left.\frac{\beta_I}{2N^2}{a\surd\sigma}\right|_a
=
\left.\frac{\surd\sigma}{g^2N}\right|_a
=
\left.\frac{\surd\sigma}{g^2N}\right|_{a=0}
+
\frac{c}{\beta_I}
\label{eqn_Kgcont}
\end{equation}
as displayed in Fig.~\ref{fig_Kg_so6}. Note that we have used the
mean-field improved coupling, $\beta_I = \beta \bar{u}_p$, which is
commonly used to improve the approach to the continuum limit
\cite{G_D3N}.
Taking the SU(4) value from
\cite{G_D3N}
we find 
\begin{eqnarray}
{\surd\sigma}/{g^2} & = & 0.4365(19)  \qquad SO(6) \nonumber\\
{\surd\sigma_{2A}}/{g^2} & = & 0.8832(41)  \qquad SU(4)
\label{eqn_KgSO6SU4}
\end{eqnarray}
which implies that 
\begin{equation}
\left.g^2\right|_{so6} =    2.023(13) \left.g^2\right|_{su4}
\label{eqn_gSO6SU4}
\end{equation}
which is again consistent with eqn(\ref{eqn_gso6}) within 2
standard deviations.

These calculations not only serve to demonstrate the equivalence 
of SO(6) and SU(4) gauge theories at the nonperturbative level where
the differing global nature of these groups might have played some
role, but they also give us confidence that calculations in SO($N$)
gauge theories hold no hidden problems.

\subsection{the deconfining transition}
\label{subsection_Tc}

SO($2N$) gauge theories should deconfine at some temperature
$T=T_c = O(\surd\sigma)$ just like SU($N$) gauge theories and
we expect deconfinement to coincide with the spontaneous
breaking of the $Z_2$ symmetry. That is to say, one can locate the
deconfining transition just as one does for SU($N$), see e.g.
\cite{Tc_D3N,Tc_D4N}.

\begin{figure}[htb]
\begin	{center}
\leavevmode
\input	{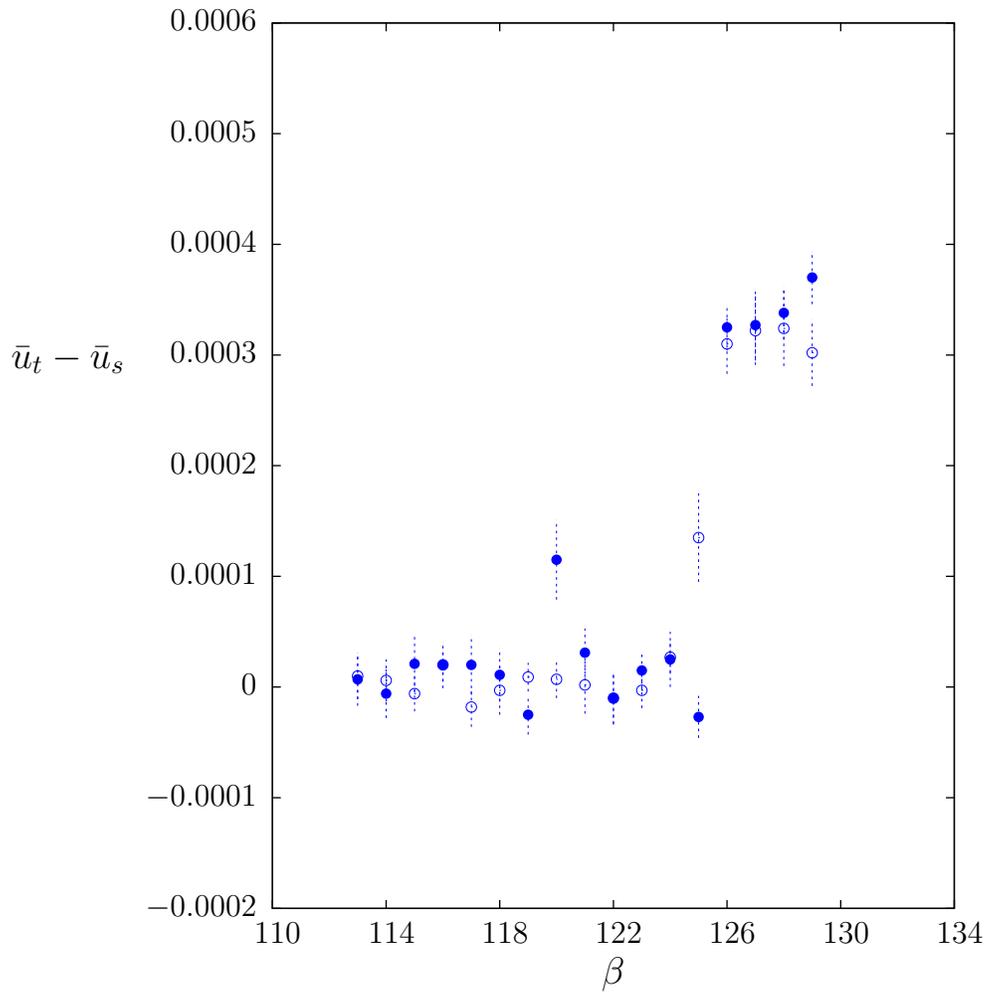}
\end	{center}
\caption{Difference between average spatial and temporal plaquettes on
a $20^2 5$ lattice in SO(12) as $\beta$ is reduced, $\bullet$, and
then increased, $\circ$.}
\label{fig_lc5n12}
\end{figure}

To illustrate the transition we consider a $20^2 5$ lattice in SU(12).
In the relevant range of couplings this spatial volume turns out to be 
large and so we can consider it to be at a well defined temperature 
$T=1/5a(\beta)$. By varying $\beta$ we vary $a(\beta)$ and hence $T$.
In Fig~\ref{fig_lc5n12} we show the value of the difference between
the average spatial and temporal plaquette as we first decrease $\beta$
and then increase it. At large $N$, volume independence tells us that
this quantity should be zero in the confining phase, and so in that limit
it acts as an exact order parameter. In  Fig~\ref{fig_lc5n12} we see
a clear transition at $\beta \sim 125$. To locate the transition 
on a spatial volume $V$ one can form a `susceptibility' from this plaquette 
difference, calculate its value at several neighbouring values of $\beta$, 
interpolate using reweighting, and define the transition $\beta_c(V)$ to be
the maximum of this suceptibility. One can now repeat this for various $V$ 
and extrapolate $\beta_c(V)$ to $\beta_c(\infty)$. This standard strategy,
see e.g. 
\cite{Tc_D4N},
can provide very precise values of the critical coupling. However
just locating the transition region from scans such as that plotted in 
Fig~\ref{fig_lc5n12} provides a value of $\beta_c$ that is accurate enough 
for our purposes in this exploratory study. In principle one can also
attempt to identify the order of the transition by looking for hysteresis
effects, but we do not attempt to do so here.

\begin{figure}[htb]
\begin	{center}
\leavevmode
\input	{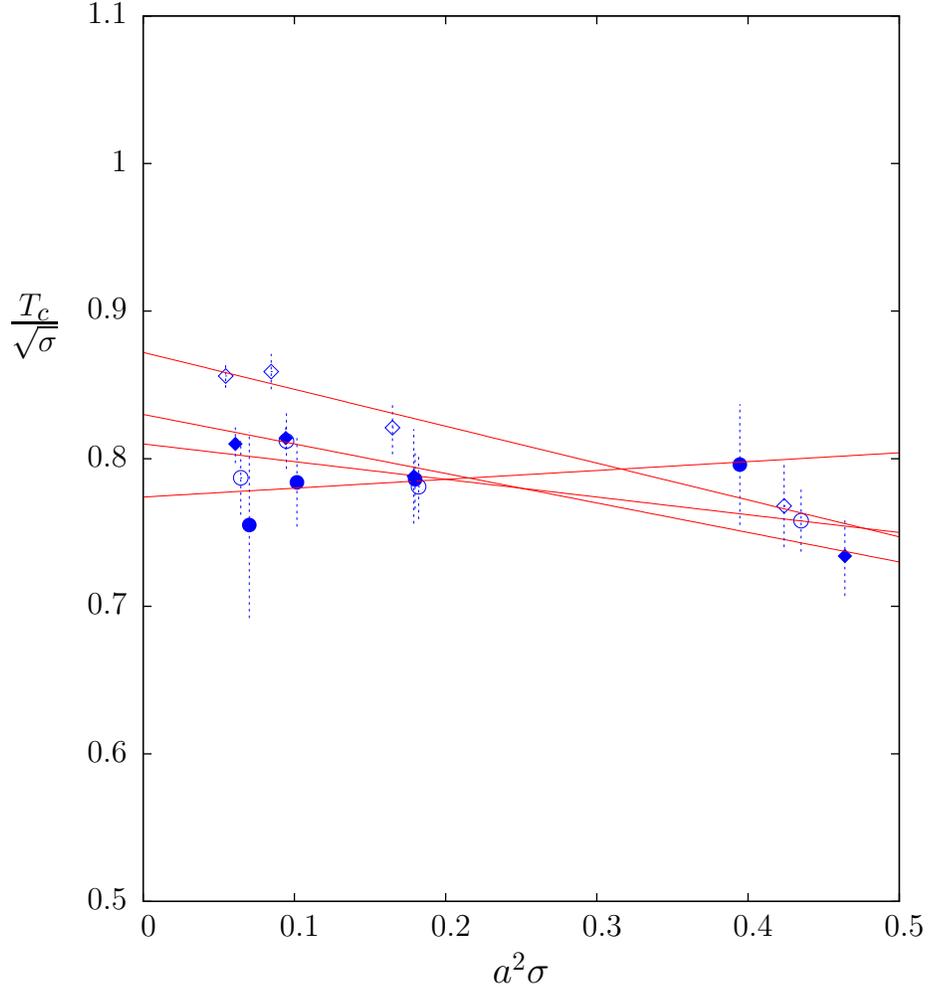}
\end	{center}
\caption{Lattice values of the deconfining temperature $T_c$ in units
  of the string tension with continuum extrapolations shown. 
For SO(4), $\bullet$, SO(6), $\circ$, SO(8), $\blacklozenge$, and SO(12), $\diamond$.}
\label{fig_aTc}
\end{figure}

We have performed such scans for $L_s^2 L_t$ lattices with $L_t=2,3,4,5$
and typically for 2 or 3 values of $L_s$ in each case to check that
finite $V$ corrections are negligible at our level of accuracy. We have 
simultaneously calculated the string tension at the resulting values of $\beta_c$
to give us an estimate of $T_c/\surd\sigma=1/\{a(\beta_c)\sqrt{\sigma}L_t\}$.
(The calculations of $a^2\sigma$ have been performed on lattices with
$L_s\surd\sigma \in [2.5,4.0]$ and $L_t>L_s$, so that they are
effectively at $T=0$.) These values are listed in Table~\ref{table_betac}.
We can then extrapolate to the continuum limit using a leading $O(a^2)$ correction 
\begin{equation}
\left.\frac{T_c}{\surd\sigma}\right|_a
=
\left.\frac{T_c}{\surd\sigma}\right|_{a=0}
+
c a^2\sigma
\label{eqn_TcKcont}
\end{equation}
We have done this for the SO(4), SO(6), SO(8) and SO(12) gauge theories,
and the results, with continuum extrapolations are shown in Fig.~\ref{fig_aTc}.
The resulting continuum values are listed in Table~\ref{table_Tc}. Here we
also show the known value for SU(4)
\cite{Tc_D3N}
(using the $k=2A$ string tension) and we observe that the SO(6) and SU(4)
values of $T_c/\surd\sigma$ are entirely consistent. 

\begin{figure}[htb]
\begin	{center}
\leavevmode
\input	{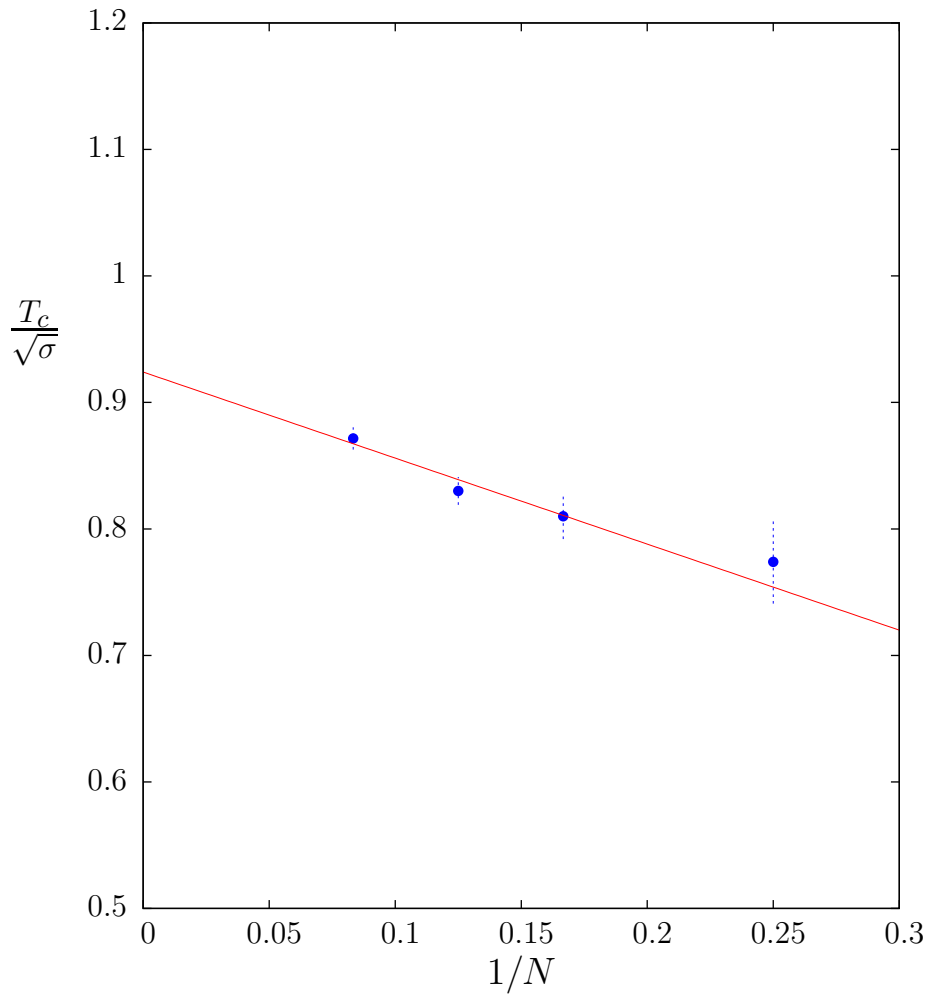}
\end	{center}
\caption{Continuum values of the D=2+1 SO($N$) deconfining temperature $T_c$ in units
  of the string tension plotted against $1/N$ with a large-$N$
  extrapolation shown.}
\label{fig_TcN}
\end{figure}

Finally we extrapolate our results to $N=\infty$ using a leading $O(1/N)$ 
correction, as shown in Fig.~\ref{fig_TcN}. This gives us the $N=\infty$
value displayed in Table~\ref{table_Tc}. We list there the SU($\infty$) value
\cite{TcD3N}
which we can see is consistent with the  SO($\infty$) value, hence providing a
confirmation of the large-$N$ equivalence of SU($N$) and SO($N$) gauge theories.

\subsection{continuum mass ratios}
\label{subsection_fixed_a}

Our above calculations of $T_c/\surd\sigma$ required us to calculate string tensions
at values of $\beta$ close to $a(\beta)=1/L_tT_c$, with $L_t\in [2,5]$. We calculated
glueball masses at the same time, and we will now use these calculations to estimate
the continuum limit of various dimensionless physical ratios 
just as we did earlier for SO(6). Of course the difference with the latter 
calculation is that the range of $a$ used for the continuum limit is much
smaller now. In fact the bulk `transition', $\beta_b$,  more-or-less coincides with
the  deconfining transition, $\beta_c(L_t)$, on a lattice with $L_t=3$. Thus, we are not
surprised to  find that we cannot perform statistically credible continuum extrapolations
with  weak-coupling corrections if we include the masses
at $\beta \simeq \beta_c(L_t=2)$. However we find that extrapolations
are often possible from a value of $\beta \simeq \beta_c(L_t=3)$,
i.e. from the bulk transition region onwards into weak coupling. Some
evidence that we are not being too optimistic is given by 
our results for SO(6) where we have performed calculations to much weaker couplings.
In Fig.~\ref{fig_MK_so6}, we see that
the extrapolations to the continuum of typical mass ratios pass
through the $\beta_c(L_t=3)$ values but not through the values at $\beta_c(L_t=2)$.
For $\surd\sigma/g^2$ we see in Fig.~\ref{fig_Kg_so6} some deviation
even from the $\beta_c(L_t=3)$ values, but it is not large. So while
some of our extrapolations do have a mediocre $\chi^2$, most are good,
and we can expect the overall picture to be qualitatively reliable.

\begin{figure}[htb]
\begin	{center}
\leavevmode
\input	{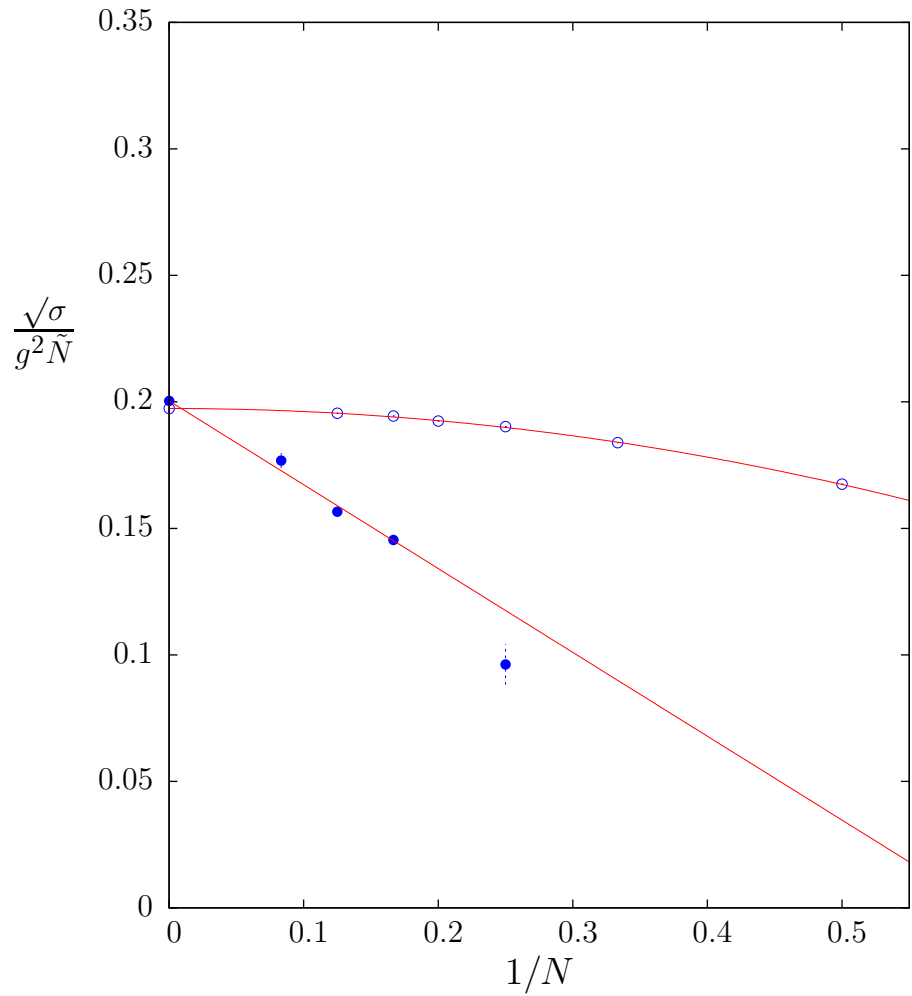}
\end	{center}
\caption{The SO($N$) continuum string tension,  $\bullet$, in units of 
 the  't Hooft coupling modified by using $\tilde{N}=N/2$,
  with an $O(1/N)$ extrapolation to $N=\infty$ shown. Also
 shown are known  SU($N$) values, $\circ$, in units of the standard 't
 Hooft coupling, $\tilde{N}=N$, and with an $O(1/N^2)$ extrapolation of
 these to $N=\infty$. }
\label{fig_KgN}
\end{figure}

In  Fig.~\ref{fig_KgN} we display our SO($N$) continuum values for the string
tension in units of the 't Hooft coupling, $g^2N$, modifed so that
$N\to N/2$ i.e. we double the calculated values of $\surd\sigma/g^2N$.
 We observe that the values
for $N\geq 6$ can be extrapolated to $N=\infty$ with just the leading
$O(1/N)$ correction. For comparison we have shown the SU($N$)
values, with an unmodifed 't Hooft coupling and we show an $O(1/N^2)$
fit to these. We observe that the $N=\infty$ extrapolations for
SO($N$) and SU($N$) are consistent with each other. The various 
(unmodifed) continuum string tensions for SO($N$) are listed in 
Table~\ref{table_McontN} as are the continuum extrapolations. 
Taking into account our other results, this tests the large $N$
prediction in eqn(\ref{eqn_g}) for the relationship between the SO($N$) and SU($N$)
couplings to an accuracy of $\sim\pm 2\%$.

\begin{figure}[htb]
\begin	{center}
\leavevmode
\input	{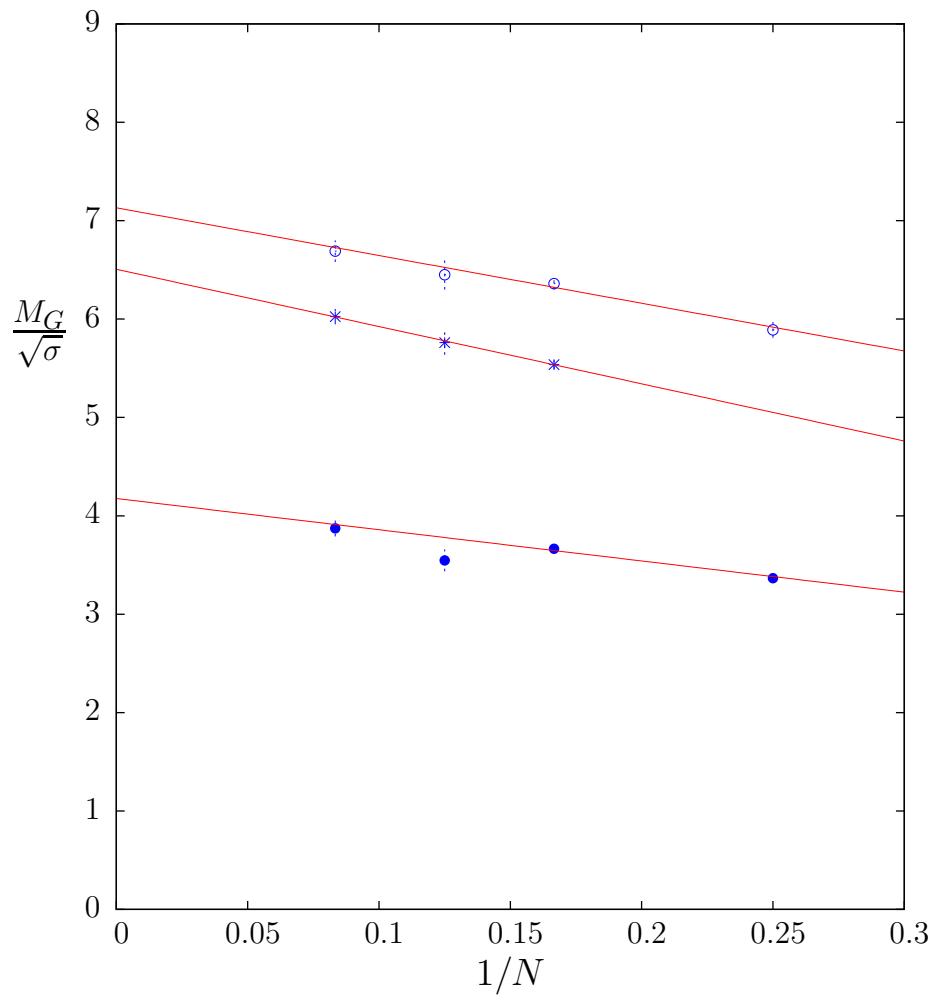}
\end	{center}
\caption{Some SO($N$) continuum glueball masses in units of the string tension: the
 lightest $0^+$, $\bullet$, and $2^+$, $\circ$ and the first
 excited  $0^+$, $\ast$. Large $N$ extrapolations shown.}
\label{fig_MKN}
\end{figure}

In Fig.~\ref{fig_MKN} we display our SO($N$) continuum values for some
of the lightest glueball masses,  in units of the string tension. We
also show the large $N$ extrapolations. All these values are listed in
Table~\ref{table_McontN} where we also list the corresponding large
$N$ limits for SU($N$) gauge theories. We see a satisfactory agreement
at the 2 standard deviation level. 

These results confirm, albeit with a modest accuracy, all our
expectations for the relationship  between SU($N$) and SO($N$) 
gauge theories in 2+1 dimensions.

\section{D=3+1}
\label{section_D4}

Finally we turn briefly to SO($N$) gauge theories in 3+1
dimensions. 

We begin with the strong-to-weak coupling bulk transition which is
easy to identify in $D=3+1$ as it is a strong first-order transition in which the
average plaquette undergoes a large and sharp dicontinuity even on
very small lattices. We have performed calculations where we gradually
decrease $\beta$ through the transition and then, well after that
transition, we gradually increase $\beta$. Because the transition is
strongly first order we have a substantial hysteresis effect, and the
two locations of the 
transitions obtained in this way  do not coincide. We list in
Table~\ref{table_bulkd4} the bulk transitions obtained for
various SO($N$) groups. These calculations have been mostly obtained
on small $4^4$ lattices, but several checks on larger volumes show
that any finite volume corrections are small.

Any continuum extrapolation can only use values of $\beta$ on
the weak-coupling side of the bulk transition. To calculate the
continuum physics of the confining phase, the lattice size $L_s$ must be
large enough i.e. $aL_s > 1/T_c$ at the very least. If the lattice
spacing on the weak-coupling side of $\beta_b$ is very small
this may require prohibitively large values of $L_s$. Indeed it has
long been known that this is the case in SO(3). One can of course 
improve one's chances by using the strong 
hysteresis to perform weak-coupling calculations at the largest
possible value of $\beta$, i.e. $\beta_b^\downarrow + \epsilon$.

This is the value of $\beta$ at which we have performed some
test weak-coupling runs. As expected our test run in SO(3)
at $\beta=2.52$ on a `large' $32^4$ lattice reveals that we are in
a small-volume phase. The same is true in SO(4) on a  $32^4$
lattice at $\beta=4.75$ where, in addition, we observe the
spontaneous breaking of the $Z_2$ center symmetry. In SO(6)
neither of the $32^4$ or $24^3 32$ lattices appear to be clearly large 
volume. However in SO(8) we appear to have what looks like
the desired confining phase on a $24^3 32$  lattice at $\beta=20$,
although not on  a $16^3 24$ lattice. Here we appear to have 
$a\surd\sigma \simeq 0.16$. Finally in SO(16) we find that we can
obtain `large-volume' physics on a $12^3 16$ lattice at $\beta=83.5$,
where we find $a\surd\sigma \simeq 0.31$. Here we are beginning to
approach the corresponding values found in SU($N$) gauge theories at larger $N$.
Thus at our larger values of $N$ one could imagine reducing $a$ by a
further factor of $\sim 2$ or 3 so as to have a useful range of $a$ for a
continuum extrapolation. However one would not want to be calculating
dynamical fermionic properties at such large $N$. We are therefore
focusing on improving the action rather than pursuing further
calculations with the standard plaquette action.   
 
Although we are not yet in a position to compare values calculated
within SO(6) with known results for SU(3),
we can do so indirectly by predicting the SO(6) physics from the known
SU(4) physics
\cite{G_D3N}. 
Doing so we have the comparison in Table~\ref{table_d4_SO6SU3}. 
We observe that the physics is very similar except where it involves
the string tension, and this is simply because the SO(6) string
tension corresponds to the $k=2A$ SU(4) string tension. 

Finally we comment that, just as for $D=2+1$, 
we can make use of the SO(6)-SU(4) equivalence
to predict the physics in $D=3+1$ 
of all $SO(N\geq 6)$ theories if we assume that
the leading $O(1/N)$ correction dominates and that SO($N$) and
SU($N$) gauge theories have a common large-$N$ limit.

\section{Conclusions}
\label{section_conclusion} 

Our aim in this paper has been to outline what we know about
SO($N$) gauge theories, complemented with some exploratory lattice
calculations. This is intended to serve as a useful background for more
detailed and precise numerical calculations.  Such studies, comparing
SO($N$) and SU($N$) pure gauge theories, will provide a starting point for
attempts to evade the  finite chemical potential sign problem in QCD 
using the (orbifold) large-$N$ equivalence of SO($2N$) and SU($N$)
theories
\cite{AC_mu,AC_N}.

We have seen in $D=2+1$ that the equivalence of SO(6) and SU(4) 
Lie algebras does indeed appear  to translate into an equivalence 
of the spectra -- with the string tension of the former corresponding
to the lightest $k=2$ (antisymmetric) tension in the latter. Since 
`glueball' mass ratios are very similar in SU(3) and SU(4) gauge theories
this implies that dimensionless mass ratios will also be
similar in SU(3) and SO(6) gauge theories, except where they involve
the string tension or the coupling $g^2$ (which has dimensions of 
mass in $D=2+1$), where the differences can be predicted. 

If, furthermore, one assumes that the leading $O(1/N)$ correction
dominates down to $N=6$, then one can use the SO(6)/SU(4) 
equivalence  and the known SU(4) spectrum
to predict the spectrum of SO($N$) gauge theories for
all $N\geq 6$. (One can additionally use the SO(3)/SU(2) equivalence
to pin down both the $1/N$ and  $1/N^2$ corrections with appropriate
assumptions.) Our exploratory $D=2+1$ calculations of the deconfining
temperature, $T_c$, the lightest two $J^P=0^+$ glueball masses
and the lightest  $J^P=2^+$ glueball mass, indicate that the $O(1/N)$ 
correction does indeed dominate all the way down to $N=6$ and often
down to $N=4$.

This SO(6)/SU(4) spectral equivalence should also  hold in $D=3+1$
and suggests a similarly strong similarity between SO(6) and SU(3)
and similar predictions for $N\geq 6$ assuming the dominance of
the $O(1/N)$ correction down to $N=6$.
Here our lattice calculations have been obstructed by the first-order  
strong-weak coupling `bulk'  transition. For low $N$ this is so located that
to calculate physics on the weak coupling side would require 
extremely large lattices. Although we find that for larger $N$,
e.g. SO(16), this is no longer the case, and the location of the 
transition in physical units is not much different from that in SU(8),
what we would really like to do is
to access the continuum physics of lower $N$ SO($N$) gauge theories,
and to that end we are investigating improved actions where the
`improvement'  desired is to push the bulk transition to stronger coupling.

\section*{Acknowledgements}

Our interest in this project was originally motivated by Aleksey
Cherman,  and we are very grateful for his encouragement and useful advice.
The project originated in a number of discussions between two of the authors 
(FB and MT) and Aleksey Cherman during the 2011 Workshop on 
`Large-N Gauge Theories'  at the Galileo Galilei Institute in Florence,
and we are grateful to the Institute for providing such an ideal environment 
within which to begin collaborations.
The numerical computations were mostly carried out on EPSRC and Oxford funded 
computers in Oxford Theoretical Physics.

\clearpage

\begin{table}[h]
\begin{center}
\begin{tabular}{|c|cc|}\hline
\multicolumn{3}{|c|}{SO($N$) \ , \ D=2+1} \\ \hline
  G    & $(ag^2N)_b \sim$ & $a\surd\sigma \sim$ \\ \hline
 SO(3)  &  3.0 &  0.164 \\
 SO(4)  &  3.44 - 3.76 &  0.265 - 0.308 \\
 SO(6)  &  3.74 - 4.06 &  0.364 - 0.428 \\
 SO(8)  &  3.85 &  0.423 \\\hline
 SU($\infty$)  & $\sim 2.2$  &  $\sim 0.7$ \\\hline
 \end{tabular}
\caption{Critical 't Hooft coupling for the D=2+1 bulk transition.}
\label{table_bulk}
\end{center}
\end{table}

\begin{table}[h]
\begin{center}
\begin{tabular}{|c|c|c|c|c|c|c|}\hline
\multicolumn{7}{|c|}{SO(6) \ , \ D=2+1} \\ \hline
  $\beta$   &  lattice  & plaq  &  $a\surd\sigma$ &   $aM_{0^+}$ &   
   $aM_{0^{+\star}}$ &      $aM_{2^+}$ \\  \hline
15.0   &  $4^2 12$  & 0.54264  & 0.678(3)  &  1.92(4)  &  --  &  -- \\
15.15 & $6^2 12$  &  0.55271  & 0.667(12) & 1.872(28) & -- & -- \\
17.75 & $8^2 16$ & 0.66735   & 0.4279(9) &  1.521(11) & -- & 2.65(13) \\
18.0  &   $8^2 16$ & 0.67383 & 0.4161(14) & 1.482(19) & -- & 2.72(19) \\
21.75 & $12^2 24$ & 0.74255 & 0.3100(7)  & 1,1277(70) & 1.696(23) &
1.89(4) \\
22.0 & $12^2 16$ & 0.74599 & 0.3053(8) & 1.1122(52) & 1.643(16) &
1.940(28) \\
25.5 & $16^2 40$ & 0.78553 & 0.2512(7) & 0.9206(46) & 1.391(10) &
1.597(13) \\ 
29.0 & $12^2 48$ & 0.81410 & 0.2108(5) & 0.7517(42) & 1.087(26) &
1.299 (47) \\
29.0 & $16^2 48$ & 0.81410 & 0.2150(4) & 0.7872(42) & 1.152(32) &
1.394(18) \\
29.0 & $20^2 48$ & 0.81410 & 0.2153(4) & 0.7852(45) & 1.183(27) &
1.386(12) \\
 29.0 & $24^2 32$ & 0.81410 & 0.2155(6) & 0.7770(51) & 1.212(13) &
1.405(14) \\
 36.0 & $24^2 32$ & 0.85294 & 0.1661(7) & 0.6011(67) & 0.947(7) &
1.065(25) \\
 48.0 & $32^2 40$ & 0.89152 & 0.1205(4) & 0.4489(36) & 0.6667(72) &
0.7709(93) \\  \hline
 \end{tabular}
\caption{Our D=2+1 SO(6) calculations with parameters and some calculated
  quantities.}
\label{table_SO6}
\end{center}
\end{table}

\begin{table}[h]
\begin{center}
\begin{tabular}{|c|cc|c|}\hline
\multicolumn{4}{|c|}{$\mu/\surd{\tilde{\sigma}}$ \ , \ D=2+1} \\ \hline
\multicolumn{1}{|c|}{$\mu$} & \multicolumn{2}{|c|}{$SO(6)$} &
\multicolumn{1}{|c|}{$SU(4)$} \\
       & $O(a^2)$ & $O(a^4)$ & \\ \hline
  $M_{0^+}$              &  3.675(27)   &  3.723(47)    &  3.638(25) \\
  $M_{0^{+\star}}$    &   5.66(7)      & 5.52(10)       &  5.48(5) \\
  $M_{2+}$               &  6.47(8)        &  6.39(14)        &  6.16(8) \\
  $T_c$                    &  0.810(18)    &      &  0.817(5) \\  \hline
 \end{tabular}
\caption{Continuum limit of some glueball masses and the deconfining
  temperature  in D=2+1 SO(6) and SU(4) gauge theories, all in units of the
  string tension $\tilde{\sigma}$. In some cases we show  $O(a^4)$ as well as $O(a^2)$
  extrapolations. In SU(4) $\tilde{\sigma}$ is the k=2 antisymmetric string tension.} 
\label{table_E_SO6SU4}
\end{center}
\end{table}

\begin{table}[h]
\begin{center}
\begin{tabular}{|c|cc|cc|cc|cc|}\hline
\multicolumn{1}{|c}{} & \multicolumn{2}{c|}{SO(4)} &
\multicolumn{2}{|c|}{SO(6)} & \multicolumn{2}{|c|}{SO(8)} & \multicolumn{2}{|c|}{SO(12)} \\ \hline
$l_t$ & $\beta_c$ & $a\surd\sigma$ &  $\beta_c$ & $a\surd\sigma$ &
$\beta_c$ & $a\surd\sigma$ & $\beta_c$ & $a\surd\sigma$ \\ \hline
 2  &  6.475(99) &  0.628(32)  &  15.15(15) &  0.660(18) &  27.35(50)
 & 0.681(25)  &  63.75(75) &   0.651(24) \\
 3  &  7.45(10)   &  0.424(11)  &  17.75(25) &  0.427(12) &  33.25(75)
 &  0.423(17) &  81.25(75) &   0.406(9) \\
 4  &  8.30(20)   &  0.319(12)  &  21.75(75) &  0.308(7)   & 41.50(25)
 &  0.307(3)   &  102.00(75)  &  0.291(4) \\
 5  &  9.30(40)   &  0.265(22)  &  25.20(60) &  0.254(7)   &
 49.75(75)  & 0.247(4) &  125.0(10) &  0.2337(20) \\ \hline
 \end{tabular}
\caption{Values of $\beta$ at which  D=2+1 SO($N$) theories reach a temperature $T=1/al_t$
  at which they deconfine. The corresponding value of the string tension,
  $a^2\sigma$, is listed.} 
\label{table_betac}
\end{center}
\end{table}

\begin{table}[h]
\begin{center}
\begin{tabular}{|c|c||c|c|}\hline
\multicolumn{2}{|c||}{SO($N$)} & \multicolumn{2}{||c|}{SU($N$)} \\ \hline
$N$ & $T_c/\surd\sigma$ & $N$ & $T_c/\surd\sigma$ \\ \hline
 4    &  0.774(33)    &      &   \\
 6    &  0.810(18)    &  4  &  0.817(3) \\
 8    &  0.830(11)    &      &  \\
 12  &  0.8715(88)  &      &  \\  \hline
$\infty$ & 0.924(20) &  $\infty$ & 0.903(23)  \\  \hline
 \end{tabular}
\caption{Continuum limit of deconfining temperature in units of the
  string tensions for various SO($N$) gauge theories and the
  $N\to\infty$ extrapolation. For comparison we show the SU(4) and
SU($\infty$) values from \cite{JLMT}.}
\label{table_Tc}
\end{center}
\end{table}

\begin{table}[h]
\begin{center}
\begin{tabular}{|c|c|c|c|c|}\hline
\multicolumn{5}{|c|}{SO($N$) \ , \ D=2+1}\\ \hline
$N$ & $\surd\sigma/g^2N$ & $M_{0^+}/\surd\sigma$   & $M_{2^+}/\surd\sigma$ & $M_{0^{+\star}}/\surd\sigma$ \\ \hline
 4    &  0.0481(40)    &  3.366(33)    &  5.89(8)      &   -- \\
 6    &  0.0727(4)      &  3.665(21)    &   6.36(6)     &   5.537(43) \\
 8    &  0.0783(8)      &  3.547(111)  &   6.45(15)   &   5.761(125) \\
 12  &  0.0884(15)    &  3.873(82)    &   6.69(11)   &   6.025(78) \\  \hline
SO($\infty$)  & 0.1002(23) &  4.18(8)   &  7.13(14)  &  6.51(16)  \\  
SU( $\infty$) & 0.0988(2 )$\times 2$ &  4.11(2)   &  6.88(6)    &  6.21(5)    \\  \hline
 \end{tabular}
\caption{Continuum limits of various mass ratios for various SO($N$) gauge theories and the
$N\to\infty$ extrapolations. For comparison we show the  known
SU($\infty$) values.}
\label{table_McontN}
\end{center}
\end{table}

\begin{table}[h]
\begin{center}
\begin{tabular}{|c|cc|}\hline
\multicolumn{3}{|c|}{SO($N$) \ , \ D=3+1}\\ \hline
   G    & $\beta_b \downarrow$ &  $\beta_b \uparrow$ \\ \hline
 SO(3)  &   2.48(1)   & 2.53(1)  \\
 SO(4)  &   4.62(3)   &  4.87(3)  \\
 SO(5)  &   7.35(5)   &  7.95(5) \\
 SO(6)  &   10.85(5)  & 11.8(1)  \\
 SO(7)  &   14.77(8)  & 16.58(8) \\
 SO(8)  &   19.7(1)    & 21.9(1) \\
 SO(9)  &   25.12(12)   &  28.12(12)\\
 SO(10)  &   31.12(12)  &  35.12(12)\\
 SO(16)  &   82.25(25)  &  93.25(25) \\ \hline
 \end{tabular}
\caption{Values of $\beta$ at the bulk transition in various D=3+1
  SO($N$) gauge theories obtained mainly on $4^4$ lattices.
 Separately for $\beta$ decreasing and increasing.}
\label{table_bulkd4}
\end{center}
\end{table}

\begin{table}[h]
\begin{center}
\begin{tabular}{|c|c|c|}\hline
\multicolumn{3}{|c|}{D=3+1} \\ \hline
      &  $SO(6)$  &  $SU(3)$  \\ \hline
    $M_{2^+}/M_{0^+}$           &   1.45(5)   &   1.35(4) \\
    $T_c /M_{0^+}$                 &    2.13(5)  &   2.29(5)  \\
    $M_{0^+}/\surd\sigma$   &    2.87(6)  &   3.55(7)  \\  \hline
  \end{tabular}
\caption{Comparing some continuum energy ratios in  D=3+1 between SU(3) and
that expected for SO(6) from its equivalence with SU(4). (Using the $k=2A$
string tension.)}
\label{table_d4_SO6SU3}
\end{center}
\end{table}

\clearpage

\clearpage

\end{document}